\documentclass[graybox, secnum]{svmult}
 
\usepackage{mathptmx}       
\usepackage{helvet}         
\usepackage{courier}        
\usepackage{type1cm}        
\def\beq{\begin{equation}}

\def\eeq{\end{equation}}
\def\beqa{\begin{eqnarray}}
\def\eeqa{\end{eqnarray}}

\def\bea{\begin{eqnarray}}
\def\eea{\end{eqnarray}}

\usepackage{varioref,exscale,latexsym,amsmath,amssymb}
\usepackage{graphicx}
\usepackage{slashed}
\usepackage{dcolumn}
\usepackage{bm}
\usepackage{hyperref}
\usepackage{color}
\usepackage{xcolor}

\usepackage{makeidx}         
\usepackage{graphicx}        
\usepackage{multicol}        
\usepackage[bottom]{footmisc}
\usepackage{hyperref}        
\usepackage{soul}            
\hypersetup{colorlinks=true,urlcolor=blue}

\usepackage{hyperref}
\hypersetup{
colorlinks,
linkcolor={black},
citecolor={blue},
urlcolor={red}}

\usepackage[square,numbers]{natbib}
\usepackage{amsmath}
\makeindex             

\begin{document}
\title*{Quantum General Relativity and Effective Field Theory}
\author{John F. Donoghue \thanks{corresponding author} }
\institute{John F. Donoghue, Department of Physics, University of Massachusetts, Amherst, Amherst, MA 01003 , \email{donoghue@umass.edu}}
%
%
\maketitle
\abstract{This is a review of some of the concepts and results of the effective field theory treatment of quantum General Relativity. Included are lessons 
 of low energy quantum gravity, and a discussion of the limits of effective field theory techniques.}

\section*{Keywords} 
Effective Field Theory, General Relativity, Quantum Gravity, Quantum Field Theory

\section{Introduction}

Perhaps without realizing it, we have lived through a paradigm change in the way that we understand the fundamental interactions. Historically we started out uncovering what we call classical physics, and then found awkward ways to describe quantum physics. The quantum techniques themselves have evolved and are now different than even 30 years ago.  The logic of effective field theory now permeates the field. We also now give primacy to the quantum theory, and have some modest understanding of how the classical world emerges from the quantum world.  This shift in our understanding is particularly important in the case of General Relativity.  The earliest quantum techniques did not work well for General Relativity. However, the modern viewpoint is well-suited for gravity. We have a theory of quantum General Relativity which treats it, within various limits, as an effective field theory. This brief review is a chance to reflect on the basic ideas and lessons of this effective field theory. 

 Our fundamental theory is now defined by a path integral over the dynamical  degrees of freedom  guided by a local Lagrangian.  Physics is an experimental science and through much effort we have determined the particles and the structure of their interactions. In a compact notation (and hiding about 26 parameters - masses, couplings, etc.), our core theory is presently given by
\beq
Z^{core} =\int [d\phi d\psi d A dg]_\Lambda~e^{i\int d^4x \left[-\frac14 F^2 +{\bar \psi} i\slashed{D}\psi +\frac12 \partial \phi \partial \phi -V(\phi) -\Gamma {\bar \psi} \phi \psi  -\Lambda_{cc} +\frac{2}{\kappa^2} R +...\right] }\ \ .
\eeq
 We use the path integral because canonical quantization is not a useful way to treat the gauge interactions of the Standard Model, while path integrals are simple and direct. Here the subscript $\Lambda$ is meant to indicate that the path integral is to be restricted to those energies which we have experimentally explored - i.e. those below some scale $\Lambda$. We do not, and need not, pretend that these degrees of freedom are alway the correct ones. The ellipses indicate that we expect further local terms in the action, suppressed by heavy mass scales, even if we have not identified them yet. That is, we think of all of our core theory as an effective field theory valid at ordinary energies.

From this starting point, General Relativity is also fundamentally a quantum theory. The metric degrees in freedom need to be dynamical and they need to be included  in the path integral because otherwise we could not obtain the classical physics such as gravitational waves.  

The idea of effective field theory is that the low energy degrees of freedom organize themselves as quantum fields, governed by a local 
Lagrangian, in general containing so-called nonrenormalizable terms suppressed by powers of a heavy scale. Nevertheless, one can make predictions without knowledge of the full high energy theory.

The basic theme of this exposition is that General Relativity is a quantum field theory which is not much different than the other interactions in our core theory. It is explored using the tools of effective field theory. In the region of validity of the effective field theory, it can be studied in perturbation theory and the quantum corrections are small - it is the best perturbative theory ever. However the lessons learned are nevertheless interesting. 

In other reviews, I have presented more of the technical details of the general relativistic effective field theory (GREFT) \cite{Donoghue:1994dn, Donoghue:1995cz, Donoghue:2012zc, Donoghue:2017ovt, Donoghue:2017pgk}, see also  \cite{Burgess:2003jk,Burgess:2020tbq}, as well as other expositions on effective field theory \cite{Weinberg:2016kyd, Weinberg:2021exr, Petrov:2016azi, Meissner}   Here I attempt a broader overview.

\section{Does the graviton exist?}

 The reader who finds this question annoying or boring is free to skip this section. However I include it because it bears on the rationale for effective field  theory.  
 
One often hears the question of whether the graviton exists. Some people will never be satisfied until they see the clicks in a detector caused by a graviton. With this criterium, there will not be a resolution in our lifetimes, or perhaps ever. But for some the question means to ask whether it would be consistent to have everything but gravity be described by quantum fields, but to have gravity be classical. As we presently understand quantum theory, this is not possible. The following brief comments illustrate how the existence of quantum fields is required. 

Steven Weinberg in particular has presented the argument that any quantum theory satisfying Lorentz invariance, causality, crossing symmetry and cluster decomposition will be described by a quantum field theory \cite{Weinberg:2016kyd, Weinberg:2021exr} This is part of the reasoning that all of our theories are effective field theories. Let us see how this could be applied to the gravitational interaction\footnote{ This form of the argument comes from D. Carney \cite{Carney:2021vvt} }. The Newtonian potential has the Fourier transform 
\beq
-\frac1{\bf{q}^2} =- \frac1{(\bf{p}_1-\bf{p}_2)^2}    \ \ .
\eeq
But for this to be consistent with special relativity, this must be made into a Lorentz invariant.
\beq
-\frac1{(\bf{p}_1-\bf{p}_2)^2} \to  \frac1{(p_1-p_2)_\mu (p_1-p_2)^\mu} =\frac1{(p_1-p_2)^2}  =\frac1{q^2}\ \ .
\eeq
We now need to specify how we deal with the pole in this function. Of course we know that the correct answer is the Feynman propagator
\beq
\frac1{q^2 +i\epsilon}
\eeq
and it is this which emerges from the path integral treatments. However, if we are trying to be more general,
Carney  \cite{Carney:2021vvt} has shown that alternate prescriptions with retarded or advanced Green functions, i.e. with $(q_0\pm i\epsilon)^2 -{\bf q}^2$ does not satisfy unitarity, nor do these forms satisfy causality. The propagator using $-i\epsilon$ is equivalent to the one with $+i\epsilon$, although the arrow of causality is reversed \cite{Donoghue:2019ecz, Donoghue:2020mdd}. But now we are basically done. Because of the identity
\beq
\frac1{q^2 +i\epsilon} = P\frac1{q^2} -i\pi \delta (q^2)
\eeq
we see that what originally was the potential comes accompanied by massless on-shell radiation - the graviton.  This combination yields graviton exchange and graviton emission. The general principles of relativity, unitarity and causality have turned the potential into a quantum propagator.

This conclusion can also be addressed in different ways. As a recent example, the authors of Ref. \cite{Belenchia:2018szb, Danielson:2021egj} show that in order to resolve a gedanken experiment involving superposed charges or masses, in electromagnetism or gravity, requires the existence of radiation - photons or gravitons. There are several previous arguments which say that for consistency, the gravitational potential must also be accompanied by gravitons  \cite{Page:1981aj, Carlip:2008zf, Giampaolo:2018yql}.

We have also partially addressed this issue using the path integral mentioned in the introduction. The metric appears in the Lagrangian because we need to describe particles in curved space. With the rest of the interactions described by the path integral, could we leave the graviton out of the integration variables? If we did, we would not obtain the graviton propagator. This in turn would not allow the interaction of two masses, which occurs due to one graviton exchange. With the integration over the gravitational field we obtain Einstein's equation as the equation of motion as well as the causal graviton propagator\footnote{Using $e^{-iS}$ in the path integral instead of $e^{iS}$ results in the time reversed propagator with the $-i\epsilon$ prescription \cite{Donoghue:2019ecz}}. In Section \ref{lessons}, we will see explicit examples of how further classical physics emerges from the path integral. So our starting point for the other fields also points to the quantum nature of the graviton. It is worthwhile to note although we often refer to the classical limit as $\hslash \to 0$, in fact $\hslash$ is a fixed number (here often set equal to unity) and classical physics emerges in the appropriate kinematic regions. It is not that we have classical physics and then treat quantization as an optional extra step. Rather the modern view is that our starting point is quantum and that the understanding of the classical limit is the extra step.

\section{Detour into QED}

  The effective field theory treatment is not a change in quantum field theory. It is rather using regular QFT with a careful attention to the scales in the problem. The common features of EFT are also seen in other theories in certain limits. In other publications, I have used the sigma model to motivate EFT techniques \cite{Donoghue:1995cz, Donoghue:2012zc, Donoghue:2017ovt, Donoghue:2017pgk}, and indeed that analogy is very useful for gravity. In order to here give a different example,  we can use QED for this purpose.
  
The QED path integral is given by  
\beq\label{QED}
Z[J] = \int [dA_\mu d\psi]_\Lambda  ~e^{i\int d^4x [{\cal L}_{QED}(A,\psi)-J_\mu A^\mu]}
\eeq
The subscript $\Lambda$, implying a limited range of this theory, is also appropriate here. This is because we know that the photon is not the correct degree of freedom at all energies. Above the scale of electroweak symmetry breaking, it is replaced by linear combinations of the $ SU(2)_L$ gauge field $W^3_\mu$, the hypercharge field $B_\mu$ and the Higgs boson. Treating the photon as a separate field is only valid below the electroweak scale. But we do not need to know that fact for QED to work at low energies. Similar comments apply to the charged fermion here. At high energy, the fermion mass eigenstates are decomposed into different fields (the weak eigenstates) and also involve the Higgs field.  

We can explore this at even lower energies - below the mass of the fermion. In this case, in situations where the external fermions are not present, they still propagate in loops in the original theory but can be removed from the effective field theory for photons. That is, we can form an effective action for the photons by integrating out the massive fermion. 
We do this by performing the path integral over the fermion field to define an effective Lagrangian involving only the photon
\beq
 e^{i\int d^4x {\cal L}_{eff}(A)}=   \int [d\psi]_\Lambda e^{i\int d^4x [{\cal L}_{QED}(A,\psi)} \ \ .
 \eeq
This leaves behind the effective field theory defined by
\beq
Z[J] = \int [dA_\mu ]_m  ~e^{i\int d^4x [{\cal L}_{eff}(A)-J_\mu A^\mu]}  \ \ .
\eeq
The subscript on the path integration is now the mass $m$ rather than the electroweak scale, because the effective field theory is only valid below that mass. 

 In practice we can do this in perturbation theory by matching the full theory to the effective theory. At leading order in the electric charge, this involves the vacuum polarization diagram, $\Pi_{\mu\nu}i(q)= \left(q_\mu q_\nu -\eta_{\mu\nu}q^2 \right) {\Pi}(q^2) $ which is described in momentum space using dimensional regularization
 \bea\label{vacpol}
 { \Pi}(q^2) &=&   \frac{\alpha}{3\pi}  \left[\frac1{\epsilon} +\log 4\pi-\gamma- \log \frac{m^2}{\mu^2} - \frac{q^2}{5m^2} \right] ~~~~~~~(q^2<<m^2) \nonumber \\
 &=&
    \frac{\alpha}{3 \pi}  \left[\frac1{\epsilon} +\log 4\pi-\gamma- \log \frac{-q^2}{\mu^2} \right]~~~~~~~(q^2>>m^2)  
\eea    
where $\epsilon =(d-4)/2$. At low energy only the top line is relevant. We know what to do with the divergence - it goes into the renormalization of the electric charge. However we should also note the $\log m^2$ term. This is present even for the heaviest masses. However, if we are to measure the electric charge at $q^2=0$, it also goes into the definition of the charge. The residual describes the deviation from the result at $q^2=0$,
 \bea
 {\hat \Pi}_{\mu\nu}(q) =  { \Pi}_{\mu\nu}(q) -  { \Pi}_{\mu\nu}(0) =\frac{\alpha}{15\pi}  \left(q_\mu q_\nu -\eta_{\mu\nu}q^2 \right) \frac{q^2}{m^2}  ~~~~~~~(q^2<<m^2)   \ \ .
\eea        
This result can be described by an effective Lagrangian containing only the photon field\footnote{Here I have chosen the $1/4e^2 $ normalization to underscore the renormalization of the electric charge is applied at $q^2=0$.}
  \beq\label{QEDeff}
 {\cal L}_{eff}(A) = -\frac{1}{4e^2(0)} F_{\mu\nu}F^{\mu\nu} - \frac{ 1}{240\pi^2 m^2} F_{\mu\nu} \Box F^{\mu\nu}  \  \ .
  \eeq
 The form of this Lagrangian has been chosen to match with the vacuum polarization amplitude when a matrix element is taken.
 
The most important point here is that the new term in the effective Lagrangian is local. This follows from the uncertainty principle. Effects from high energy/momentum appear only at short distance in coordinate space. Such effects can be Taylor expanded in the the light momenta and are then represented by a local derivative expansion for the effective Lagrangian. The takeaway is that effects from high energy appear local when viewed at low energy, and that they can be represented by local Lagrangians. 

The other important property demonstrated here is the decoupling of the heavy mass.  You can see through Eq. \ref{vacpol} that the vacuum polarization depends on the logarithm of the heavy mass. However, this is absorbed into the definition of the renormalized coupling. In this sense, the electric charge depends on the masses of charged particles, no matter how heavy. However, there is no physics in this dependence - we only use the measured value of the charge. This is the Appelquist-Carazzone theorem at work \cite{Appelquist:1974tg}. The effects of a heavy particle appears either in the renormalization of the coupling constants of the theory, or are suppressed by powers of the heavy mass.\footnote{An exception is when integrating out the heavy particle violates the symmetry of the theory.}.

This is perhaps a good place to note that despite our early discussion which emphasized that we treat the path integral as correct below some scale $\Lambda$, we generally do not use a cutoff in calculations. Dimensional regularization respects the symmetries of many theories where a cutoff often does not, and it is easy to use. However the loop integration does run over all scales including those beyond the applicability of the effective field theory. This is nevertheless acceptable. The "wrong" behavior at high energy appears as a local effect and satisfies the Appelquist-Carazzone theorem. It then disappears into the renormalization and identification of the parameters of the local effective Lagrangian. Those parameters will be the appropriate ones as long as the EFT is not applied outside of its range of validity. 

Another feature that can be seen here is that there is no Ostrogradsky instability \cite{Ostrogradsky, Woodard:2015zca} associated with the higher derivatives. This refers to the result in classical mechanics where theories with higher time derivatives the Hamiltonian, calculated by canonical methods, exhibits an instability. While the second term has the higher time derivatives, in practice it does not lead to any instability. At low energy its effect is small compared to the usual energy. The higher derivatives can become comparable, and potentially trigger an instability, only at energies which are far higher than those appropriate for the effective field theory  \cite{Simon:1990jn}. Despite the extra derivatives, the classical limit of the effective field theory is usual E\&M. 

There are corrections to this result suppressed by more powers of the mass. The next term in the effective Lagrangian is that of Euler and Heisenberg
\beq
{\cal L}_{EH} = \frac{\alpha^2}{90m^4}\left[ (F_{\mu\nu} F^{\mu\nu} )^2+ \frac74 (F_{\mu\nu} \tilde{F}^{\mu\nu})^2\right]
\eeq
which occurs due to the box diagram. This mediates interactions of photons. 

It is also instructive to look at the opposite extreme, where the mass of the fermion goes to zero or the relative momentum transfer is large compared to the mass. In this case there can be no expansion in inverse powers of the mass. In the vacuum polarization diagram, the logarithm becomes more important.  This is non-analytic and cannot be Taylor expanded in derivatives. If we try to represent it in position space it would be non-local, represented by the non-local function
\beq
\langle x|\log \Box | y\rangle  \equiv \int  \frac{d^4q}{(2\pi)^4 } e^{i q\cdot (x-y)} \log (-q^2)
\eeq
Physically, this is non-local because massless fields can propagate long distances. If we were to try to match this to an effective action, it would also have to be non-local, schematically represented by \cite{Donoghue:2015xla}
\beq\label{nlaction}
S_{light} = \int d^4x  ~-\frac14 F_{\mu\nu}\left[\frac{1}{e^2(\mu)} - \frac{ 1}{12\pi^2}  \log \frac{\Box}{\mu^2} \right]F^{\mu\nu}  \  \ ,
\eeq
with the shorthand notation
\beq
\int d^4 x A \log \Box ~B \equiv \int d^4x d^4 y ~ A(x) \langle x|\log \Box | y\rangle B(y)  \ \ .
\eeq
In Eq. \ref{nlaction} we note the appearance of the running coupling constant. However, the main point here is that massless fields yield non-analytic terms such as $\log q^2$ in momentum space, and non-localities in position space.

The effective field theory is a full quantum field theory. This can be seen by the fact that the effective action, Eq. \ref{QEDeff}, still contains the integration over the photon field. 

 \section{General Relativity as an effective field theory}

While most pedagogic treatments of General Relativity emphasize geometry and curved spacetime, it is also possible to develop it as a gauge field theory \cite{Kibble:1961ba, Utiyama:1956sy, Donoghue:2017pgk}. If we want to obtain a field theory coupled to energy and momentum, we will gauge spacetime translations, which are the corresponding symmetries. This leads to general covariance, the metric as a dynamical field and covariant derivatives. The action then must be an invariant, with the simplest terms being the cosmological constant ant the Einstein action. The geometric treatment is exceptionally powerful for the classical theory. The field theory treatment is conceptually closer to the development of the Standard Model, and is more useful for the quantum theory. Effective field theory then helps make sense of the quantum field theory.

The ultimate origin of the gravitational interactions may not be known, but we do know the symmetry of the theory - general covariance. The unknown physics from high energy will produce a local Lagrangian at low energy. The curvatures are second order in derivatives of the metric, so the action can be ordered in the derivative/energy expansion, with the first several terms being
\beq
S_{grav} = \int d^4 x \sqrt{-g} \left[    - \Lambda +\frac{2}{\kappa^2}R + c_1 R^2 + c_2 R_{\mu\nu}R^{\mu\nu} +...      \right]
\eeq
with $\Lambda$ being the cosmological constant and $\kappa^2 =32\pi G$. The effective field theory by itself says nothing about the magnitude of the various constants. If we do not know the underlying theory we need to measure these parameters. At ordinary energies, the effects of the curvature squared terms are negligible if the constants $c_1, c_2$ have any normal size\footnote{In particular there would only be noticeable effects in the gravitational interaction at a millimeter if the coefficients were greater than about $10^{65}$. If the derivative expansion of General Relativity is scaled by the Planck mass, we would expect these dimensionless coefficients to be of order unity.. But, given the unexpectedly small value of the cosmological constant and the possibility of new physics as we increase the energy, we should be open to the possibility that this latter expectation is not correct.}. Moreover these do not trigger an Ostrogradsky instability when used as effective interactions \cite{Simon:1990jn}  in the same way that we saw in the QED example above.

Including extra terms in the action beyond the Einstein term  is not a big deal by itself. But doing this allows one to perform renormalization of the quantum theory. By using the most general Lagrangian consistent with general covariance, we can be sure that all the UV divergences can be renormalized into the various coefficients as long as we use a regularization which does not break general covariance.  The quantum field is the fluctuation in the metric which deviates from a given background metric $g_{\mu\nu}= {\bar g}_{\mu\nu} + \kappa h_{\mu\nu}$. There is a residual covariance associated with the background field. The divergences are at short distance, which by the equivalence principle can be treated as almost flat, so that we know the behavior of propagators at short distance.  The one-loop renormalization was carried out beautifully by 't Hooft and Veltman \cite{tHooft:1974toh}. The divergences due to graviton loops can be represented by an effective Lagrangian of the form
\beq
{\cal L}_{div} =\frac1{16\pi^2} \frac1{\epsilon}\left(\frac1{120} R^2 + \frac{7}{20} R_{\mu\nu}R^{\mu\nu}\right)
\eeq
while those for other matter fields are similar but with different coefficients. One can see that these divergences can be absorbed into the renormalized values of the coefficients $c_1,~c_2$. Power counting in powers of $G$ \cite{Donoghue:2017pgk} reveals that higher order graviton loops yield divergences at higher order in the derivative expansion, i.e two loops divergences are of order $R^3$ \cite{Goroff:1985th}. In contrast, higher order loops of matter fields from renormalizable field theories remain at order $R^2$. This also can be seen by power counting, because the divergences in such theories do not have any inverse powers of the mass needed to compensate for extra powers of the derivatives in the numerator. 

However, the renormalization of divergences is also not that big of a deal, although it was the focus of this subject for many years. The divergences themselves come from the high energy end of the theory, which we know is not reliable. The ultimate UV completion will eventually tell us the correct way to treat this domain, and will predict the value of the coefficients. So renormalization is a necessary step, but one without much content.

The real power of the effective field theory is that it shifts our attention from the UV (where we do not know the physics) to the IR (where we do). There, EFT techniques allow one to make real predictions. This is because we know the light degrees of freedom active there and we know their interactions. At a given order in the energy expansion, we have a  small number of coefficients, such as $\Lambda,~ G, ~c_1,~c_2$, so that we have reduced our ignorance of the full theory of quantum gravity to a few constants. However, there are dynamical effects which are independent of these coefficients. This comes from the fact that massless fields like the graviton can propagate large distances, so that this propagation is distinct from any term in a local Lagrangian. 

As an example, let us display what happens with the gravitational interaction of non-relativistic particles, which will be discussed with more specificity in the next section. At tree level, one graviton exchange gives an amplitude which behaves as $1/q^2$ much like photon exchange in QED. When Fourier transformed this gives the Newtonian potential. At one loop level the amplitude picks up non-analytic behavior from the loops of gravitons. Schematically at the next order in $G$, we see that this has the form
\beq
{\cal M} \sim \frac{ GMm }{q^2}\left[ 1  + a G \sqrt{-m^2  q^2} +b G q^2 \log(-q^2)+  c Gq^2 +....\right]
\eeq
with $a, ~b,~c$ being constants to be calculated. The analytic term $c Gq^2$ will in practice contain effects from the coefficients $c_1, c_2$, which come from the extra derivatives when we consider the squares of curvatures in the local action. However the non-analytic terms in this matrix element will be independent of these parameters. If we are to Fourier transform the matrix element to obtain a position-space potential, it will have the form
\beq
V(r) = -\frac{GMm}{r} \left[ 1+ a' \frac{GM}{r} +b' \frac{G\hslash}{r^2} + c' G\delta^3 (x) \right]  \ \ .
\eeq
The power-law corrections come from the non-analytic terms in momentum space and are independent of the coefficients $c_1, c_2$. When we restore powers of $\hslash$ we can see that the first correction $\sim GM/r$ is classical, and the second one $\sim G\hslash/r^2$ is a quantum correction. 

The form of the low energy dynamics is similar to what happens in other quantum field theories. One does not need to know what happens at extremely high energies in order to make predictions at ordinary energies - this is the basic message of effective field theory. Knowledge of the low energy particles and interactions are sufficient. General Relativity fits beautifully into this paradigm. 

\section{Lessons of Quantum General Relativity}\label{lessons}

Perturbative quantum field theory is best at calculating transitions and scattering amplitudes. From these we can learn some features about low energy quantum gravity. We have waited a long time for a quantum theory of General Relativity. While the EFT does not answer all of our questions, let us see what we can do with it. 

\subsection{There is a universal quantum correction to the non-relativistic potential} 

We can calculate the gravitational scattering amplitude for two massive particles at one loop order. Because all diagrams are included, this is the complete quantum amplitude and it is a gauge invariant. While the full amplitude is a function of all the kinematic variables, we can take the non-relativistic limit in which the only important variable is the three momentum transfer $-q^2 = \mathbf{q}^2$. For display purposes, one can then Fourier transform this function to obtain a position space potential. As explained in the previous section, the power law corrections in $r$ follow from the non-analytic momentum dependence and are independent of any divergences or unknown coefficients. The result for two particles of mass $M$ and $m$ is \cite{Bjerrum-Bohr:2002gqz, Khriplovich:2002bt}
\beq\label{potential}
V(r)  = -\frac{GMm}{r} \left[ 1+ 3\frac{G(M+m)}{r} +\frac{41}{10\pi}\frac{G\hslash}{r^2}  \right]   \ \ .
\eeq
It is obviously the third term which is the quantum correction. 

It is interesting that result has now been calculated using three methods. The original calculations used the usual Feynman diagram methods. However, the same result can be obtained by modern unitarity-based methods .\cite{Bjerrum-Bohr:2013bxa, Holstein:2016cyq}
Here only the on-shell gravitational Compton amplitudes (those involving two on-shell gravitons) are required. These are related to the unitarity cut in the crossed channel. By evaluating this cut and mapping it onto the cuts of the  master Feynman integrals, one can obtain the final result more simply. In addition, because of the property that the graviton amplitude is related to the square of a gauge theory amplitude (the gauge-gravity correspondence or ``double copy'' \cite{Bern:2002kj, Bern:2019prr}) one really only has to evaluate the QED Compton amplitude in this way of calculating \cite{Choi:1994ax}. This avoids needing to use the very messy triple-graviton vertex. Because in this method only on-shell physical gravitons are used, there are no Fadeev-Popov ghosts involved. The original Feynman diagram calculation was done in harmonic (deDonder) gauge, while the on-shell results use a form of an axial gauge, confirming the gauge invariance of the result. The result has also been obtained via dispersion relations \cite{Bjerrum-Bohr:2013bxa}, where the spectral function also is calculated from the cut, in both harmonic gauge and using the double copy axial gauge. 

Another interesting feature of this result is that it is universal. The universality was first found by Holstein and Ross \cite{Holstein:2008sx}
by redoing the Feynman diagram calculations for particles of different spins. In such a method the universality is remarkable because different Feynman diagrams are involved in the various cases, and the universality is only seen when adding all diagrams together. However, by the unitarity and dispersive methods the result can be understood to be the consequence of tree-level soft theorems. Electromagnetic Compton amplitudes and gravitational Compton amplitudes are universal in the limit of small momenta \cite{Low:1954kd, Weinberg:1965nx, Gross:1968in}. This applies not only to elementary particles but to composite macroscopic objects.  In the unitary and dispersive methods, the non-analytic terms arise from multiplying together the universal tree amplitudes. The result is then a one-loop soft theorem for quantum gravity.

\subsection{Both classical and quantum effects come from loop diagrams}

We are often told that the loop expansion is an expansion in $\hslash$. If this were the case, we would not expect to obtain the classical correction seen in Eq. \ref{potential} from one loop Feynman diagrams. However the folk theorem is in fact not true. For the gravitational interaction, this has been known since the work of Ishikawa \cite{Iwasaki:1971vb} and Gupta and Radford \cite{Gupta:1980zu}, but the insight is more general \cite{Holstein:2004dn} . At a technical level, when one is counting powers of $\hslash$ by pulling out overall factors of this quantity, there are residual factors left behind. For example in the Dirac equation one has $\hslash~ \bar{\psi}(i \slashed{\partial} -m/\hslash)\psi$. The $m/\hslash$ factor can compensate an overall factor of $\hslash$, as in $G \sqrt{q^2 m^2} = G\hslash \sqrt{q^2m^2/\hslash^2}$. On a more philosophical level, if we are to reconstruct the world, including the classical limit, from the path integral treatment, then the classical results need to be contained somewhere in the Feynman diagram expansion. 

This insight was first developed into a calculational program for classical gravitational wave physics by Goldberger and Rothstein \cite{Goldberger:2004jt}. With some further developments it has now become a subfield for obtaining classical results from QFT techniques. For recent reviews, see \cite{Bjerrum-Bohr:2022blt, Bjerrum-Bohr:2022ows, Goldberger:2022rqf}

\subsection{There is no ``test mass'' limit for quantum effects}

It is common to imagine a test mass of vanishing size moving along a geodesic without itself distorting the spacetime. Perhaps surprisingly this does not work for quantum effects. The quantum effects sample more than just the geodesic even for small masses. The test mass limit for the classical interaction can be seen in the non-relativistic potential, Eq. \ref{potential}, where the classical correction depends on $M+m$. If $m$ is the smaller mass of the two, then sending $m$ to zero leaves just the larger mass determining the correction. But the quantum effect is independent of the relative masses, and both masses contribute equally.

\begin{figure}[htb]
\begin{center}
\includegraphics[height=80mm,width=140mm]{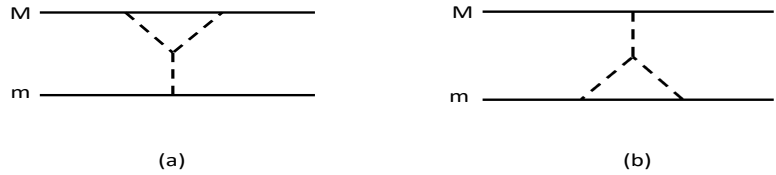}
\caption{Two diagrams which contribute to the gravitational scattering of two masses.  The solid lines represent particles with a large mass $M$ and a small mass $m$. The dashed lines are gravitons. }
\label{figure}
\end{center}
\end{figure}

To see this in more detail, let us look at two of the diagrams which contribute to the corrections of the Newtonian potential, shown in Fig. 1. These diagrams contribute to both classical and quantum effects. For the classical portion, Fig. 1(a) gives a correction which is proportional to $M$ and that of Fig.1(b) yields one  proportional to $m$. So in the test mass limit Fig. 1(b) is negligible, and it is reasonable to interpret that of Fig. 1(a) as a correction to the metric that the test body moves in. However, for the quantum effects, both diagrams 1(a) and 1(b) give equal quantum corrections, so the idea of a test mass does not work for these diagrams. In Fig. 1(b), the graviton propagates for a long distance and samples the gravitational field not only along the geodesic but also at different points. It is an irreducible tidal effect. As a correction to the Newtonian potential, it does not vanish as the mass is taken to zero. There are other diagrams with this property also.

\subsection{The bending of massless particles is not universal}

One can address the bending of light also by a scattering calculation. This should not be done by calculating the quantum cross-section and then using the classical relation to the bending angle - that procedure only works for $1/r$ potentials. However, the eikonal approximation is designed to recover the geometric optics result at large impact parameter $b$. By calculating the peak of eikonal phase one reproduces the classical bending angle to impressively high order in inverse powers of the impact parameter \cite{Akhoury:2013yua, Bjerrum-Bohr:2016hpa}. One can add quantum corrections to this result. The use of untiarity methods, where the diagrams are reconstructed from the on-shell cuts, simplifies the calculation greatly. 

As usual, the result is unobservably small at any reasonable impact parameter. However the interesting aspect is that it is not the same for all massless particles \cite{Bjerrum-Bohr:2014zsa, Bai:2016ivl, Chi:2019owc}. One finds
\beq\label{angle}
\theta = \frac{4GM}{b}+ \frac{15}{4}\frac{G^2M^2\pi}{b^2} + \frac{8c_b -47+64\log (2r_0/b)}{\pi} \frac{G^2M\hslash}{b^3}
\eeq
where $c_b= (371/120, ~113/120,~-29/8 )$ for scalars,  photons and gravitons respectively\footnote{The original calculation only included gravitons in the cuts. Subsequent calculations  \cite{Bai:2016ivl, Chi:2019owc}. correctly also include massless matter particles.} and $r_0$ is an infrared cutoff. We should not be surprised at this lack of universality, as there are no low energy theorems for massless particles as there were in the non-relativistic limit. The tidal effects include the long distance propagation of gravitons and the massless particles themselves. There are entirely different diagrams involved for the different cases. 

This result has interesting implications for the concepts of light cones and geodesics, as will be discussed below.

\subsection{$G$ and $\Lambda$ are not running couplings in physical processes}

We are used to having our coupling constants depend on the energy scale at which one measures it. The idea of a running coupling captures the effects of quantum processes relevant for that scale. These are universal because they come along with the renormalization of the couplings. In a mass independent scheme, the $1/\epsilon$ in dimensional regularization are always accompanied by $\log ({\rm Energy})^2/\mu^2$, where $({\rm Energy})$ represents some of the kinematic variable in the process under investigation \footnote{The use of a mass independent scheme is useful to avoid confusion with $1/\epsilon - \log m^2/\mu^2$ which does not indicate kinematic running.}. 

In the gravitational action the coefficients quadratic in the curvature, i.e. $c_1,~c_2$, obey this paradigm. We have seen that loops of massless particles, including gravitons, renormalize the coefficients of operators of order $R^2$. This can be seen most simply in the gravitational vacuum polarization diagram, and here the divergences do come along with factors $\log q^2/\mu^2$. In analogy with the electromagnetic case in Eq. \ref{nlaction} , we can represent this physical effect to an action such as
\beq\label{BV}
\sim \int d^4x \sqrt{-g} \left[c_1(\mu) R^2  + b R \log(\Box/\mu^2) R +...\right]  \ \ .
\eeq 
This is exactly what is done in the formalism developed by Barvinsky and Vilkovisky  \cite{Barvinsky:1985an, Barvinsky:1995jv, Barvinsky:1993en, Satz:2010uu}. It is straightforward to convert this into a renormalization group equation for $c_1$. 

But $\Lambda$ and $G$ are not like this. When we calculate loops of gravitons or other massless particles using dimensional regularization, $\Lambda$ and $G$ are not renormalized - only the curvature squared terms are. So the logarithms which accompany renormalization are not present. 

However, there is renormalization of $\Lambda$ and $G$ when one has loops of massive particles, although these have a different character. For example for a scalar of mass $m$ the one loop divergence relevant for the cosmological constant is \cite{Donoghue:2022chi}
\beq\label{deltaLambda}
\delta \Lambda = -\frac{m^4}{32\pi^2} \left[  \frac1{\epsilon} -\gamma +\log(4\pi) +\log \frac{\mu^2}{m^2} +\frac32 \right] \ \ . 
\eeq
Note that the logarithm here is $\log m^2$. It has nothing to do with the external scales of the problem. Once one measures the cosmological constant at one scale, it does not change when working at a different scale. Following the $\log \mu$ dependence does not signify running in physical reactions when masses are present because $\log m^2/\mu^2$ does not change with scale. Another example of this sort  is in the QED vacuum polarization for $q^2<<m^2$ , i.e. Eq \ref{vacpol}. At low energy the $\log m_t^2/\mu^2 $ from a top quark in the loop does not lead to running. For the reader who would like to see this realized in a EFT setting close to gravity, but with real comparison to experiment, can study the renormalization of the pion decay constant $F_\pi^2$ in chiral perturbation theory  \cite{Gasser:1984gg, Donoghue:1992dd}, which plays the same role in the chiral EFT as $1/G$ does in General Relativity. 

One can also make this theoretical argument by noting that the running of these parameters with external scales would need to match on to some effective action, most likely non-local. For logarithmic running we have seen now how this works at the curvature squared order in gravity, Eq. \ref{BV}, and in QED, Eq. \ref{nlaction}.  For the cosmological constant and for the Einstein term, general covariance says that there are no non-local operators which share the form of the local operator. The leading non-local operator closest to the cosmological constant has been calculated \cite{Donoghue:2022chi}, but it has a different structure. In particular in an expansion about flat space, the cosmological constant has a term linear in the gravitational field, while any non-local partner must have at least two fields. 

When using a cutoff regularization, one finds divergences depending on powers of the cutoff such as ${\rm (cutoff)}^4$ or ${\rm (cutoff)}^2$ for $\Lambda$ and $G$\footnote{However, some of the naive dependences often quoted are in fact absent due to cutoff dependent terms in the path integral Jacobean \cite{Donoghue:2020hoh}.}. It needs to be emphasized that these do not define running parameters in physical processes. On one hand, physical results do not depend on regularization scheme, and the absence of these dependences in dimensional regularization implies that they are not physical. They are absorbed when measuring $\Lambda$ or $G$ at a given scale and do not change when the external scales do. In an effective action framework, powers of external scales would be represented in the local Lagrangian by powers of derivatives or curvatures. They are already included in the local terms in the derivative expansion of the EFT, but with fixed coefficients. Again, the related case of chiral perturbation theory is an experimentally tested EFT where one verifies these comments.

Finally, one could just try to identify a running coupling in the calculation of physical processes. Perhaps there turns out to be some useful way to capture the quantum effects as a function of scale. There are multiple calculations which have been done. In any one process, one could identify some energy dependence which describes the quantum effects for that process. But there is no useful or universal identification that works with multiple processes \cite{Anber:2011ut} 

The subfield of Asymptotic Safety \cite{Percacci:2007sz, Niedermaier:2006wt, Reuter:2019byg} uses a IR cutoff and defines the theory by using Wilsonian ideas for running that cutoff from a UV fixed point down to zero energy. However, this is just used to define the physical theory once the full range of the cutoff is included. It does not mean that the physical parameters run, and is not intrinsically in conflict with the above discussion. But it should be recognized that it is incorrect to use  that power-law cutoff dependence in physical settings as if it were running in phenomenological applications \cite{Donoghue:2019clr}. More modern treatments use the derivative expansion to describe energy behavior in physical processes \cite{Knorr:2022dsx}, and are more in line with the effective field theory treatment.

\subsection{There is no ``quantum metric''}

It is tempting to look for quantum effects modifying the metric describing various classical solutions. We do not do this for QED (there is no ``quantum corrected electric field'' surrounding a charge) but classical solutions play such a foundational role in General Relativity that we are certainly interested in this question. 

For cases where gravity remains classical and the quantum effects are due to matter fields, such as those involving photons in the Reissner-Nordstrom metric, this question appears well defined \cite{Donoghue:2001qc}. For example, the quantum correction due to photon loops in $g_{00}$  is\footnote{In this equation I have removed the factor of $\hslash$ from the usual definition of $\alpha$ so that $\alpha =e^2/4\pi$.}
\beq\label{metric}
g_{00} =  1-\frac{2GM}{r} + \frac{G\alpha}{r^2} -\frac{8}{3\pi}\frac{G\alpha \hslash}{Mr^3} \ \ .
\eeq
Since gravity is classical here, this result just follows from loop corrections to the energy momentum tensor.  Here again, loop diagrams reproduce both the leading classical correction and a quantum correction.

However, when gravity itself is treated in QFT, the result is more problematic. Part of the problem is that the metric itself is not a well defined quantum concept. Quantum physics traditionally describes transition amplitudes and the like, and the results described earlier in this section have been derived from these. 
But the traditional class of well-defined quantum objects does not include the field variable itself. 

Along these lines, there have been attempts \cite{Radkowski, Duff:1974ud, Bjerrum-Bohr:2002fji, Khriplovich:2004cx} (including one by the present author) to calculate the quantum corrections to the Schwarzschild metric. These have been criticized by Kirilin \cite{Kirilin:2006en} as not being invariant under the reparametrization of the gravitational field, i.e. the way that one choses to expand about a background metric. In QFT, is it not supposed to make a difference if we perform field redefinitions as long as the identification of the on-shell free fields is maintained. This is commonly used and is referred to as Haag's theorem \cite{Haag}. However, even in non-gravitational QFT the theorem only applies to on-shell matrix elements. Off-shell quantities and intermediate results are not similarly invariant. A given set of quantum corrections to a metric are intermediate results to a full calculation and not only depend on the gauge (which is to be expected) but also on the field parameterization (which is harder to overcome). 

The semiclassical idea of the expectation value of the metric $\langle g_{\mu\nu} \rangle$ is also not a valid quantum object when the gravitational field itself is quantum. It is subject to Kirilin's criticism. One would need to understand how this expectation value would be measured in a quantum process for it to be well defined. 

This problem as described above is one of perturbation theory. The idea of a covariant non-local effective actions seems well defined. If treated fully, this presumably could reveal the nature of quantum solutions. But in practice such actions are only approximately known. The method of Barvinsky and Vilkovisky called the expansion in the curvature \cite{Barvinsky:1985an, Barvinsky:1995jv, Barvinsky:1993en}, and referred to briefly in Eq. \ref{BV} is an example. It appears expressed in terms of curvatures, so that might be independent of field redefinitions. However, there is not a unique understanding of what is meant by $\log \Box$ - this part of the calculation is not expressed in terms of curvatures. The correction to the logarithmic term in the effective action involves structures which are generically of the form $R^2\frac1{\Box}R$. These are of the same order in both the loop expansion and in the derivative expansion as the logarithmic correction. Dimensional analysis can be used to show that in the Schwarzschild case, near the horizon these corrections are of the same magnitude as the logarithmic ones. So even here we see some limits of perturbation theory. 

In a less formal context, we can also see in the calculations above that there is not a universal quantum metric.  We can look at the calculations and see if there is a corrected metric which would describe these situations. One can easily see that there is no single metric around a massive body which would recover the behavior for both the non-relativistic and massless particles.  This result also implied indirectly in the discussion of test particles. The diagrams that go into the calculation of a change in the metric, which would include Fig. 1a, are not the final result, which would also include Fig. 1b.

\subsection{Light cones/Penrose diagrams appear as uncontrolled approximations}

It is perhaps redundant to point out that the results described above call into question many of our standard tools. If massless particles follow different trajectories, which one is it that defines the light cone? If there is an intrinsic fuzziness to this concept, how does one draw a proper Penrose diagram? To be sure, these effects are small in the limits where the EFT is valid. Using the standard tools in that regime will be approximately valid, with calculable corrections. But the effects get larger and more important as one approaches the limits of the EFT, where many of the interesting quantum gravity questions are posed. These tools are not controlled approximations outside of the EFT region.

\section{Limits of the gravitational effective field theory}

The heart of effective field theory is the idea that only the degrees of freedom which are active at a given energy need to be included and we only need to know their interactions near that scale. Physics is an experimental science and there are boundaries to what we know. Here are some comments on the limits of the effective field theory for gravity.

\subsection{High energy}

The most obvious limit to the effective field theory is at high energy, or large curvature. At some energy our knowledge of the right degrees of freedom or of their interactions  fails. We then need a more complete theory.  Or perhaps if the same ingredients remain valid, near the Planck scale we would enter a strongly coupled regime where EFT techniques would be useless. 

Many of the most interesting question in the study of quantum gravity are sensitive to the high energy limit. In EFT jargon, this is referred to as the need for a UV completion. Many of the other contributions to this volume describe these theories. 

\subsection{The extreme infrared limit of the theory}

    Effective field theory is meant to be best in the infrared, and there is no indication that this is not correct. However, there are technical limitations on what we can do with present techniques, which  become most obvious in the extreme infrared. These are perhaps more interesting than the high energy limitations, as they may lead to new techniques and perhaps new insights.
    
    The local effective action which we start with is an expansion in the local curvature. In general the nonlocal quantum effects are calculated by a perturbative expansion around a background metric, $g_{\mu\nu} =\bar{g}_{\mu\nu}+ \kappa h_{\mu\nu}$. However, gravitational effects build up and the metric can get large even when the local curvature is small. By the equivalence principle, we can alway chose the background metric to be almost flat in a neighborhood of any point. This is most clearly seen using normal coordinates, which expand about a position in spacetime
\beq
g_{\mu\nu} (x') = \eta_{\mu\nu} + \frac13 R_{\mu\alpha\mu\beta} (x) y^\alpha y^\beta+...~~~~~,~~~~~y=(x'-x) \ \ .
\eeq
However, if the distance away is far enough, the metric in this expansion will become large unless higher order terms in the expansion, of order $R^2$ and higher, become important. An extreme example of this is the standard Schwarzchild coordinates for a black hole. Coordinates which are smooth at infinity have a metric which blows up at the horizon.  If we choose coordinates which are smooth near a point on the horizon, they will blow up somewhere else. This happens even though the curvature itself will be small outside and on the horizon for massive black holes. 

This problem also permeates classical perturbation theory. One can use post-Newtonian expansions to calculate gravitational radiation for the far-field part of the inspiral phase, but we need numerical techniques to capture the results near the horizon scale even when the curvature there is small. 

However, the problem is more severe in quantum perturbation theory. We have seen that the quantum gravity effects are non-local because the massless propagators can probe large distances. The non-locality implies that even if we are in a region of small metric deviation, the propagators can probe the larger metrics further away. Most of our field theoretic techniques are adaptions from flat space methods. For a specific example, consider the logarithmic non-locality such as we have discussed frequently above. In flat space, dimensional analysis tells us that for time independent problems
\beq
\langle \mathbf{x}| \log \Box|\mathbf{y} \rangle =  \int  \frac{d^3q}{(2\pi)^3 } e^{-i \mathbf{q}\cdot (\mathbf{x}-\mathbf{y})} \log (\mathbf{q}^2) \sim \frac1{(\mathbf{x}-\mathbf{y})^3 }
\eeq
The resulting $1/r^3$ dependence is seen in the non-relativisitic potential Eq. \ref{potential}, the bending angle Eq. \ref{angle}, and  the metric correction Eq. \ref{metric}.  However, this means that if you were to use this form in the analysis of the Schwarzschild solution, even if you were at a large distance from the center, you would be sensitive to the horizon where the metric blows up and sensitive even to the curvature singularity at the origin. 

Again this seems to be a technical problem, which could potentially be solved by numerical relativity. Intuitively we expect such effects to be small - we should not need to know about the black hole at the center of our galaxy in order to do weak field calculations on earth. But we presently do not have a universal estimator for how big such corrections are. And we do not have any proof that the quantum effects do not build up over long distances like the classical effects do \footnote{Yang Mills theory also is a good perturbation theory at short distances but, for very different reasons, at large distance becomes non-perturbative. At the very largest distances, QCD and QED in the real world have neutral states so that the growth with distance is not important. Gravitational charges are all the same sign, so effects can build up.}.  So this becomes a limitation on the EFT techniques. 

Perhaps the issue could be addressed by combining patches using different coordinates and matching on the boundary. In each patch we can use the equivalence principle to make the coordinates nearly flat. Then matching at the boundaries would convey the information from one patch to another. However, this program has not yet been carried out.

\subsection{What are the right quantum questions?}

Quantum field theory in curved spacetime is a challenging subject even for non-gravitational particles and interactions. It is not even clear how to rigorously define a particle in general curved spacetimes. However, in lightly curved worlds we can approximately use Minkowski field theory techniques. We do this all the time since we live in a lightly curved spacetime. In curved spacetime, the gravitational effective field theory shares these challenges, and adds an extra one because the metric is now a dynamical variable. 

The perturbative solution is to expand around a background metric and treat the fluctuating field quantum-mechanically. We have seen that this can yield leading predictions at low energy and curvature. But the effective theory has also demonstrated that some of our usual techniques have some limitations. Some of these can be traced to the perturbative expansion, as we have seen the difficulty of identifying the backreaction on the background metric itself. 

The larger challenge is to identify valid quantum questions for gravity which are also able to be answered with techniques which we know how to apply. This may require the adoption of nonperturbative methods, at least in cases where the metric field becomes strong. Even in nonperturbative computational methods such as Causal Dynamical Triangulations, it is also a challenge to identify sensible objects to be calculated \cite{Loll:2019rdj}. The limits of perturbation theory itself may prove to be one of the limits of the effective field theory.  In any case, there still are interesting field theory developments possible even within the effective field theory.

\subsection{Other potential limits}

Effective field theory is a ``humble'' theory in that it attempts to work within the framework of existing knowledge, and to acknowledge its limits. In this regard, we should recognize that there are other limits on our understanding besides just the energy/distance variable. 

In particular, quantum mechanics itself has only been tested within some limits. In reactions, there are generally only a few particles involved. Even in condensed matter physics, with enormous numbers of particles, the interactions are primarily two-body or sometimes three-body. We presently have no experimental need for modifications to quantum mechanics, as we do for other interactions outside their known frontiers. However there could be a ``macroscopicity'' frontier. Many issues in the transition from quantum physics to classical physics and decoherence remain poorly understood. Gravity could be the place where these effects become important, as gravity is not screened and very large masses are available. Recent work on stochastic effects and decoherence, reviewed by C. Burgess in this volume \cite{Burgess}, provides some new techniques within quantum mechanics. But perhaps there could be real changes with quantum physics for objects that are macroscopic enough. This point of view has been put forward by Penrose  \cite{Penrose:1976js} and by Stamp and collaborators \cite{Stamp:2015vxa}.

\section{Conclusion}

Claims that General Relativity clashes with Quantum Mechanics are wrong, or at least misleadingly simplistic. The covariant quantization of General Relativity is by now an old topic \cite{Feynman:1963ax, DeWitt:1967ub, Donoghue:2017pgk, Buchbinder:2021wzv}. Modern quantum field theory techniques - effective field theory - help us to extract physical quantum predictions. These techniques are completely normal quantum field theory methods which are applied routinely in other settings. So the quantum field theory of quantum General Relativity is quite normal.

Indeed, General Relativity is perhaps the best example of the effective field theory paradigm. It appears to be valid over many orders of magnitude in distance or energy, and it overlaps with quantum physics in regions where the latter is relevant. We have tested the low energy degrees of freedom, and their interactions follow from a simple action. Plus, since we do not know the ultimate UV completion for quantum gravity, the effective field theory is all that we can be sure about. 

There is no reason to be misleadingly simplistic about quantum gravity. We have made strides in quantizing, renormalizing and applying quantum General Relativity. The effective field theory results are interesting in themselves. And we can readily motivate further quantum gravity work without being misleading. What we should be saying is that quantum General Relativity points unmistakably to our lack of understanding of the full theory.

Quantum gravity is not optional. While we likely will not know the experimental outcome in our lifetimes regarding the various options, there is still much to learn about the consistency and structure of the theories. Perhaps the exploration will change our understanding of the world. But we do know that all of the theories of quantum gravity need to reduce to the effective field theory of General Relativity in the appropriate limit. This effective field theory provides a foundation for our exploration of quantum gravity. 

\section*{Acknowledgement} The writing of this review has been supported in part by the US National Science Foundation grant NSF-PHY-21-12800.

\end{document}